\documentclass[pra,aps,twocolumn,showpacs,epsfig,showkeys]{revtex4-1}
\usepackage{color}
\usepackage{graphicx,amssymb,epsfig,epsf,amsmath} 

\begin{document}

\title{Destruction of attractive bosonic cloud due to high spatial coherence in tight trap} 

\author{Anindya Biswas$^{1}$, Barnali Chakrabarti$^{2,3}$, Tapan Kumar Das$^{1}$ and Luca Salasnich$^{4}$}
\affiliation{
$^{1}$ Department of Physics, University of Calcutta, 
92 A.P.C. Road, Kolkata 700009, India.\\
%\email[e-mail : ]{abc.anindya@gmail.com}
$^{2}$ Instituto de Fisica, Universidade of S\~ao Paulo, CP 66318, 05315-970, S\~ao Paulo,Brazil.\\
$^{3}$ Department of Physics, Lady Brabourne College, P1/2 Surawardi Avenue, Kolkata 700017, India.\\
$^{4}$ Dipartimento di Fisica "Galileo Galilei" and CNISM, Universita di Padova, Via Marzolo 8, 35122 Padova, Italy.
%CNR and CNISM, Unit\`a di Padova, 
%Dipartimento di Fisica ``Galileo Galilei'', Universit$\grave{a}$ 
%di Padova, Via Marzolo 8, 35122 Padova, Italy.
}

\begin{abstract}
We study coherence of trapped bosonic cloud with attractive finite-range interaction in a tight harmonic 
trap. One-body density and pair distribution function in the 
ground state for different trap sizes are calculated. We also calculate healing length and the correlation length which 
signify the presence of high spatial coherence in a very tight trap leading to the destruction of 
the condensate for a fixed particle number. This is in marked variance with the usual collapse of the 
attractive metastable condensate when $N > N_{cr}$. Thus we investigate the critical frequency and critical size of 
the trap for the existence of attractive BEC. The finite range interaction gives a nonlocal effect in the 
effective many-body potential and we observe a high-density stable branch besides the known 
metastable branch. Moreover, the new branch shows universal behavior even in the very tight trap. 
\end{abstract}



\maketitle

\section{Introduction}
Trapped ultracold bosonic atoms have attracted a revival of interest since the achievement of the Bose Einstein 
condensate of dilute gases. The standard defining property of Bose-Einstein condensation 
(BEC) is the macroscopic occupation of the 
single particle ground state for the ideal system. In the case of an interacting system, BEC is often related to the 
off-diagonal long-range order~\cite{pen,ujh}. By controlling the external magnetic field, one can virtually manipulate the 
interatomic force as well as the trap size. This can change the situation  from a weakly interacting case to a very strongly interacting one and obviously 
the role of interatomic correlations come in the picture. Experimental relevance stems from the fact that experiments  designed to produce the interference pattern in BEC 
are closely related with the study of coherence properties.\\ 
%The dimensionality of the system also plays a key role.
%If the transverse degrees of freedom are frozen, one has effectively one-%dimensional system. There are many detailed 
%calculation in the study of first and second order coherence in strongly interacting 1D bosons, which are well 
%described by the Lieb-Liniger model and also in the homogeneous Bose gas~\cite{sak,nar,ast,ner,ijy,khr,gan}. It has developed a revival of interest as the 
%direct experimental demonstration of off-diagonal long-range order in the interference experiment is possible~\cite{rit,tgg}.\\
%
%The purpose of the present work is different. 

We investigate a weakly interacting Bose gas in the 
presence of an external trap. We consider a finite number of bosons interacting via the finite range van der Waals interaction, 
instead of the commonly used zero-range contact interaction. The $-\frac{C_{6}}{r^{6}}$ tail introduces an attractive non local interaction. 
$^{7}$Li is particularly interesting due to its negative scattering length and when the number of bosons is below some threshold 
value, a metastable condensate appears. Due to the attractive interaction, the effect of correlation becomes important even in the weakly interacting gas and we expect to get new physics in the study of correlation properties.  Although it is 
commonly believed that the Gross-Pitaevskii (GP) equation, based on a mean-field approach and a contact interaction, is adequate for weakly interacting Bose gases, but in the  
present study we utilize an approximate many-body technique which is more rigorous and retains all possible two-body 
correlations~\cite{das,das1,ffd}. 
This technique, called correlated potential harmonics (CPH) method, uses 
a subset of full hyperspherical harmonics (HH) expansion basis. While the 
latter is exact and incorporates all correlations in the many-body wave 
function, CPH basis retains only two-body correlations and ignores 
higher-body correlations. This assumption is manifestly valid in a typical 
laboratory BEC, which must be extremely dilute to avoid depletion due to 
recombination through three-body collisions. The HH method is very 
cumbersome and is practicable only for three-body systems. But in the CPH 
method, the bulk of superfluous basis functions are eliminated, so that it 
can be adopted for a large number of bosons (see Sec.~II for details). It has been successfully 
applied to attractive BECs and repulsive systems containing up to 14000 particles~\cite{das1,ffd}. 
The method is important especially in the study of correlation properties in realistic condensates. Thus the motivation of our 
present work is as follows. Firstly, to study different correlation properties
 in the ground state, like one-body density and pair-distribution function for few hundred $^7$Li atoms
 in the external three-dimensional trap. This facilitates the investigation of the effect of interatomic correlations in the many-body wave function  
on the bulk correlation properties of the condensate. 
To quantify the latter, we also calculate the healing length which is another key quantity in the 
study of coherence properties. Secondly, we investigate the effect of the change of trap size on the correlation 
properties. By controlling the trap frequency it is possible to make tighter traps. We observe that in a tighter trap the metastable condensate 
becomes highly correlated. Effective correlation length and the 
healing length reduce drastically, as the trap size decreases; however, 
they remain larger than the trap size, 
%[as can be seen from Fig.~8 (see later)],  
indicating strong bulk spatial correlation within the metastable condensate. 
In addition to the metastable condensate, the many-body system exhibits a 
strongly attractive narrow well, which supports cluster states (see below). 
Below a critical trap size the metastable region in the effective 
many-body potential disappears. 
This manifests the destruction of a metastable condensate in a very tight trap. We estimate the critical trap size above which BEC can be observed. 
The disappearance of the condensate in a very tight trap significantly differs from the usual collapse of attractive 
BEC in a large trap.\\

Due to the presence of a realistic interaction with a strong repulsive core, we have 
a deep but finite attractive well outside a repulsive core on the left side of 
the metastable region in the effective potential. Hence, besides the appearance of 
the low density metastable BEC, we also find a stable branch at high density which 
leads to the formation of atomic clusters. With decreasing trap size, the 
metastable branch shrinks and eventually disappears (resulting in a collapse due 
to squeezing). 
However the attractive well of the effective potential supporting the 
high density branch remains invariant in position, magnitude and shape 
with decrease in trap size. Thus 
it exhibits a universal behavior. The energy and size of the cluster do 
not change with the trap size. 
In earlier calculations~\cite{Salasnich,Parola,Reatto}, a non-local effective interaction 
in the momentum space, which is equivalent to a local repulsive contact interaction 
plus a finite-range attractive interaction in the coordinate space was adopted in a 
variational calculation of the GP equation. This also produced the narrow 
attractive well and the associated high-density branch, in addition 
to the low-density metastable branch. On the other hand, a purely attractive local contact 
interaction in the GP equation produces a pathological essential singularity at 
the origin, besides the metastable region. In Ref.~\cite{Salasnich,Parola}, quantum 
tunneling rates between the low- and high-density branches and the decay rates due 
to two- and three-body collisions were reported. Nature of transitions between 
the two branches as a function of the number of bosons was also investigated. In 
the present study, we obtain these two branches as a result of using the 
realistic van der Waals interaction having a strong short-range repulsion (due to 
nucleus-nucleus repulsion) and a long-range attraction, in an approximate 
many-body theory, which incorporates all two-body correlations. We focus on 
the correlation properties and investigate the effect of decreasing trap size. \\

The paper is organized as follows. The many-body calculation technique is presented in Sec.II. Sec.III
presents the calculation of many-body effective potential and different correlation properties like one-body density and pair distribution function. Sec.IV contains a summary and the conclusions.\\

\section{Many-body calculation with correlated Potential harmonics basis}
The Hamiltonian for a system of $(N+1)$ atoms ( each of mass $m$) and interacting via two-body potential has the form
\begin{equation}
H = - \frac{\hbar^{2}}{2m} \sum_{i=1}^{N+1}\nabla_{i}^{2}+ \sum_{i>j=1}^{N+1}V(\vec{x}_{i}-\vec{x}_{j})+ 
\sum_{i>j=1}^{N+1}V_{trap}(\vec{x}_{i})
\end{equation}
where $V(\vec{x}_{i}-\vec{x}_{j})$ = $V(\vec{r}_{ij})$ is the two-body potential described later and $\vec{x}_{i}$ is the 
position vector of the $i$-th particle. We use the standard Jacobi 
coordinates defined as $\vec{\zeta}_{i}$ = $\left(\frac{2i}{i+1}\right)^{\frac{1}{2}}\left[ \vec{x}_{i+1}- \frac{1}{i} 
\sum_{j=1}^{i} \vec{x}_{j}\right]$, $(i=1,2,....N)$ and the center of mass through $\vec{R}$ = $\frac{1}{N+1}\sum_{i=1}^{N+1}\vec{x}_{i}$. 
Then the relative motion of the atoms is described in terms of $N$ Jacobi vectors $(\vec{\zeta}_{1}, \cdot\cdot\cdot, \vec{\zeta}_{N})$ as~\cite{das}
\begin{eqnarray}
\left[-\frac{\hbar^{2}}{m}\sum_{i=1}^{N}\nabla_{\zeta_{i}}^{2}+ V_{trap}(r)
+V(\vec{\zeta}_1, \cdot\cdot\cdot, \vec{\zeta}_N)
-E \right] \nonumber \\
\times \Psi(\vec{\zeta_{1}},\cdot\cdot\cdot,\vec{\zeta_{N}})=0, 
\end{eqnarray} 
where $V$ is the sum of all pair-wise interactions expressed in terms of the Jacobi vectors. We assume that only two-body correlations in 
the many-body wave function are important~\cite{das,das1}, This permits the 
use of the potential harmonics expansion method, in which 
the total wavefunction $\Psi$ is decomposed into two-body Faddeev components $\psi_{ij}$ for 
the $(ij)$ interacting pair. 
\begin{equation}
\Psi = \sum_{ij>i}^{N+1} \psi_{ij}. 
\end{equation}
Note that with two-body correlations alone, $\psi_{ij}$ is a 
function only of 
two-body separation ($\vec{r}_{ij}$) and a global length, called 
hyperradius ($r=\sqrt{[\sum_1^N\zeta_i^2]}$) and is independent of the coordinates of all the particles other than the 
interacting pair~\cite{das,das1}. 
$\psi_{ij}$ (symmetric under $P_{ij}$) satisfies the Faddeev equation 
\begin{equation}
(T+V_{trap}-E)\psi_{ij}(\vec{x}) = -V(\vec{r}_{ij})\sum_{k,l>k}\psi_{kl}(\vec{x}), 
\label{Fadeveq}
\end{equation}
$T$ being the total kinetic energy; operating $\sum_{i,j>i}$ on both sides of Eq.~(\ref{Fadeveq}), we get back the original 
Schr\"odinger equation. In this approach we assume that when $(ij)$ pair interacts, the rest of the bosons are inert spectators. Moreover, 
since particle labels are unimportant, we take $\vec{r}_{ij}$ as 
$\vec{\zeta}_N$. Next we define a hyperradius  
$\rho_{ij}= \left [ \sum_{k=1}^{N-1}\zeta_{k}^{2}\right] ^{\frac{1}{2}}$ 
for the remaining $(N-1)$ noninteracting bosons\cite{das,das1}, such that 
$\rho_{ij}^{2} + r_{ij}^{2} = r^{2}$, $\rho_{ij}= r \sin \phi$ and 
$r_{ij}= r \cos \phi$. In this choice the hyperspherical coordinates 
are 
\begin{equation}
(r,\Omega_{N})= (r, \phi, \vartheta, \varphi, \Omega_{N-1})
\end{equation}
where  $(\vartheta, \varphi)$ are the two spherical polar angles of the separation vector $\vec{r}_{ij}$,
$\Omega_{N-1}$ involves $(3N-4)$ variables: $2(N-1)$ polar 
angles associated with $(N-1)$ Jacobi vectors $\vec{\zeta}_{1}, \cdot\cdot\cdot, \vec{\zeta}_{N-1}$ and $(N-2)$ angles defining the relative lengths of these Jacobi 
vectors~\cite{das}. Then the Laplacian in 
$3N$-dimensional space has the form 
\begin{equation}
\nabla^{2} \equiv \sum_{i=1}^{N} \nabla _{\zeta_{i}}^{2} = \frac{\partial^{2}}{\partial r^{2}}+ 
\frac{3N-1}{r} \frac{\partial}{\partial r}+ \frac{L^{2}(\Omega_{N})}{r^{2}}, 
\end{equation}
$L^{2}(\Omega_{N})$ is the grand orbital operator in $D$ = $3N$ dimensional space. Eigenfunctions of this operator, corresponding to all possible 
sets of quantum numbers (associated with $3N-1$ degrees of freedom), constitute 
the complete hyperspherical harmonics (HH) basis. Expansion of $\Psi$ in 
this basis would lead to an exact treatment. However, the degeneracy of 
the basis (arising from different allowed values of $3N-1$ quantum 
numbers for a particular grand orbital quantum number, ${\mathcal L}$) 
increases very fast with ${\mathcal L}$ and $N$. This fact and the fact 
that calculation of matrix elements involve $(3N-1)$-dimensional 
integrals make the calculation extremely cumbersome for $N>3$. Moreover, 
imposition of symmetry also becomes very difficult as $N$ increases. 
Hence the HH expansion method (HHEM) is practicable only for 
three-body systems. A great deal of simplification is possible for the 
laboratory BEC, which is extremely dilute and only two-body correlations 
are important. One can then use a subset [called potential harmonics 
(PH) basis] of HH basis with great advantage. 
Potential harmonics for the $(ij)$-partition 
are defined as the eigenfunctions of $L^{2}(\Omega_{N})$ corresponding to zero eigenvalue of $L^{2}(\Omega_{N-1})$ .
The corresponding eigenvalue equation satisfied by 
$L^{2}(\Omega_{N})$ is~\cite{fabre}
\begin{equation}
\left[L^{2}(\Omega_{N}) + {\mathcal{L}}({\mathcal{L}}+D-2)\right] {\mathcal P}_{2K+l}^{l,m}(\Omega_{ij})=0, \hspace*{.5cm}
{\mathcal{L}}= 2K+l \cdot
\end{equation}
This new basis is a subset of the full HH set and it does not contain any function of the coordinate $\vec{\zeta}_{i}$ with 
$i<N$. It is given by a simple closed expression~\cite{fabre} 
\begin{equation}
{\mathcal P}_{2K+l}^{l,m} (\Omega_{(ij)}) =
Y_{lm}(\omega_{ij})\hspace*{.1cm} 
^{(N)} P_{2K+l}^{l,0}(\phi) {\mathcal Y}_{0}(D-3),
\end{equation}
where $Y_{lm}(\omega_{ij})$ is the spherical harmonics 
and $\omega_{ij}= (\vartheta, \varphi)$. The function 
$^{(N)}P_{2K+l}^{l,0}(\phi)$ 
is expressed in terms of Jacobi 
polynomials~\cite{Abramowitz} and ${\mathcal Y}_{0}(3N-3)$ is the HH of order zero in
the $(3N-3)$ dimensional space, spanned by $\{\vec{\zeta}_{1}, \cdots, 
\vec{\zeta}_{N-1}\}$ Jacobi vectors\cite{fabre}.   
Thus the contribution to the grand orbital quantum number comes only from the interacting pair and the $3N$ dimensional Schr\"odinger equation 
reduces effectively to a four dimensional equation. 
The relevant set of quantum numbers (associated with the hyperangles) 
are {\it only three} -- orbital $l$, azimuthal $m$ and grand orbital 
$2K+l$ for any $N$. 
This leads to a dramatic simplification of the many-body calculations. 
Besides drastic reduction of degeneracy of the basis, potential matrix 
elements involve only three-dimensional (one-dimensional for central 
forces) integrals. The physical picture is that all irrelevant degrees 
of freedom have been frozen. Using this procedure, we have solved 
BEC containing up to 14000 bosons~\cite{das1, chak1}. The method has been 
successfully applied to attractive condensates as well~\cite{ffd,debnath, 
kundu, haldar}. 
We expand $(ij)$-Faddeev component, $\psi_{ij}$, in the complete set of potential harmonics appropriate for the $(ij)$ partition: 
\begin{equation}
\psi_{ij}
=r^{-(\frac{3N-1}{2})}\sum_{K}{\mathcal P}_{2K+l}^{lm}
(\Omega_{N}^{(ij)})u_{K}^{l}(r). 
\label{Fad-comp-expn}
\end{equation}
Note that the notation has been slightly modified to include the 
superscript $(ij)$ in $\Omega_N$, to indicate that it is the full 
set of hyperangles in $D$ dimensional space, for the particular 
choice of Jacobi vector $\vec{\zeta}_N=\vec{r}_{ij}$. 
Eq.~(\ref{Fad-comp-expn}) includes two-body correlations only. 
This is perfectly justified in the context of dilute attractive Bose gas,  where the effect of two-body correlation is important and one can safely ignore the effects of higher-body correlations. Taking projection of 
Eq.~(\ref{Fadeveq}) on a particular PH, a set of coupled differential 
equations (CDE) is obtained~\cite{das,das1} 
\begin{eqnarray}
\Big[&-&\dfrac{\hbar^{2}}{m} \dfrac{d^{2}}{dr^{2}} + \dfrac{\hbar^{2}}{mr^{2}}
\{ {\cal L}({\cal L}+1) + 4K(K+\alpha+\beta+1)\} 
\nonumber 
\\
&+&V_{trap}(r)-E  \Big] U_{Kl}(r) 
\\
&+& \sum_{K^{\prime}}f_{Kl}V_{KK^{\prime}}(r)f_{K'l} U_{K^{\prime}l}(r) = 0 ,
\nonumber
\end{eqnarray}
where $U_{Kl}(r) = f_{Kl}u_{K}^{l}(r)$,  ${{\cal L}} =
l+\frac{3N-3}{2}$, $\alpha=\frac{3N-5}{2}$, $\beta=l+\frac{1}{2}$, $l$ being the orbital angular momentum contributed by the 
interacting pair and $K$ is the hyperangular momentum quantum number.
The
potential matrix element $V_{KK^{\prime}}(r)$ is given by~\cite{das,das1} 
\begin{equation}
 V_{KK^{\prime}}(r) = \int {\mathcal P}_{2K+l}^{{lm}^{*}}(\Omega^{(ij)}_{N})
V(x_{ij})
{\mathcal P}_{2K^{\prime}+l}^{lm}(\Omega^{(ij)}_{N})d\Omega_{N}^{(ij)}.
\end{equation}
The quantity $f^{2}_{Kl}$ is given by
\begin{equation}
f_{Kl}^{2} = \sum_{k,l>k} <{\mathcal P}_{2K+l}^{lm}(\Omega^{(ij)}_{N})
|{\mathcal P}_{2K+l}^{lm}
(\Omega^{(kl)}_{N})>.
\end{equation}
It is the overlap of the PH for the $(ij)$-partition 
(corresponding to only the $(ij)$-pair interacting) with the 
sum of PHs for all partitions. An expression for $f_{Kl}^2$ can be found 
in Ref~\cite{das}.
%\hspace*{3cm}
So far we have disregarded the effect of the short range correlation in the PH basis. 
In the case of a dilute Bose gas, as the energy 
of the interacting pair is extremely small, the two-body interaction is generally 
represented by the $s$-wave scattering length $a_{sc}$ alone (as in 
the mean-field GP treatment), disregarding its detailed structure. On the other hand, a 
realistic interatomic potential like the van der Waals interaction has an 
attractive $-\frac{1}{r_{ij}^{6}}$ tail 
at larger separations and a strong repulsion at short separation. The short-range 
behavior is usually represented by a hard core of radius $r_c$. 
For a given two-body potential having a finite range, $a_{sc}$ 
can be obtained from the asymptotic solution of the zero-energy two-body 
Schr\"odinger equation 
\begin{equation}
-\frac{\hbar^2}{m} \frac{1}{r_{ij}^2}\frac{d}{dr_{ij}}\left(r_{ij}^2
\frac{d\eta(r_{ij})}{dr_{ij}}\right)+V(r_{ij})\eta(r_{ij})=0. 
\end{equation}
The value of $r_c$ is obtained from 
the requirement that the calculated $a_{sc}$ has the experimental value.
The correlation function quickly attains its asymptotic form
$(C_{1}+C_{2}/r_{ij})$ for large $r_{ij}$. The asymptotic
normalization is chosen to make the wavefunction positive at large $r_{ij}$ 
and the corresponding scattering length 
is given by $a_{sc}=-\frac{C_{2}}{C_{1}}$. In the experimental BEC, the 
energy of the interacting pair is negligible compared with the depth of the 
interaction potential. Thus $\eta(r_{ij})$ 
correctly reproduces the short $r_{ij}$ behavior of $\psi_{ij}(r_{ij},r)$. 
Hence we introduce this as a short-range correlation function in the 
expansion basis and 
call it as correlated potential harmonics (CPH) basis~\cite{ffd}. 
\begin{eqnarray}
\left[{\mathcal P}_{2K+l}^{l,m} (\Omega_{(ij)})\right]_{corr} =
Y_{lm}(\omega_{ij})\hspace*{.1cm} 
^{(N)} P_{2K+l}^{l,0}(\phi) \nonumber \\
 \times {\mathcal Y}_{0}(3N-3) \eta(r_{ij}), 
\end{eqnarray}
The
correlated potential matrix element $V_{KK^{\prime}}(r)$ is now given by
\begin{eqnarray}
V_{KK^{\prime}}(r) &=& (h_{K}^{\alpha\beta}
h_{K^{\prime}}^{\alpha\beta})^{-\frac{1}{2}}
\int_{-1}^{+1} \Big\{ 
P_{K}^{\alpha 
\beta}(z)
V\left(r\sqrt{\frac{1+z}{2}}\right) 
\nonumber 
\\
&& \times P_{K^{\prime}}^{\alpha \beta}(z)\eta\left(r\sqrt{\frac{1+z}{2}}\right)
w_{l}(z) \Big\} dz .
\end{eqnarray}
Here $h_{K}^{\alpha\beta}$ and $w_{l}(z)$ are respectively the norm
and weight function of the Jacobi polynomial
$P_{K}^{\alpha \beta}(z)$~\cite{das,fabre}.\\
Note that the inclusion of the short-range correlation function, $\eta(r_{ij})$ 
makes the PH basis non-orthogonal. This introduces an overlap matrix on the 
eigenvalue side of the matrix eigen value equation. One can use the standard procedure for 
handling a non-orthogonal basis, by introducing a transformation using the eigenvalues 
of the overlap matrix to convert the eigenvalue equation into the standard diagonalization of 
a symmetric matrix. However, we have checked that $\eta(r_{ij})$ differs from a constant value 
only by a small amount in a relatively small interval. Hence to simplify the calculation, 
we project Eq.~(4) on to a particular PH, {\it viz.} ${\mathcal P}^{lm}_{2K+l}(\Omega^{(ij)}_N)$. 
The dependence of the overlap 
$<{\mathcal P}^{lm}_{2K+l}(\Omega^{(ij)}_N)|{\mathcal P}^{lm}_{2K+l}(\Omega^{(kl)}_N)\eta(r_{kl})>$ 
on the hyperradius $r$ is quite small. Disregarding 
derivatives of this overlap with respect to the hyperradius, we approximately get back Eq.~(10), 
with $V_{KK^{\prime}}(r)$ given by Eq.~(15). 

\section{Results}
\section*{A: Choice of interaction and calculation of effective potential}
We consider the $^{7}$Li condensate with $a_{sc} = -27.3$ Bohr. For 
the numerical calculation, we choose 
the oscillator units (o.u.) of length and energy, commonly used in BEC 
calculations. The 
oscillator length, $a_{ho}= \sqrt{\frac{\hbar}{m\omega}}$ 
is chosen as the unit 
of length and the oscillator energy $\hbar \omega$ as the unit of energy, 
$\omega$ being the 
trapping frequency ($\omega=2\pi\nu$). However for presenting results, 
both in tabular form and in figures, we use MKS units: meter (m) and 
Joule (J) for length and energy respectively. 
The trap size corresponding to Rice 
University experiment is $a_{ho} = 3.0 \mu$m. Our chosen potential 
is the realistic van der Waals potential, which has 
a strong repulsive core (which is chosen as a hard core of radius $r_{c}$) 
and an attractive tail at larger separations:
$V(r_{ij})$  = $\infty$  for  $r_{ij} \leq r_c$ and
 = $-\frac{C_6}{r_{ij}^6}$  for  $r_{ij}>r_c$. 
The strength ($C_6$) is known for a given type of atom. In the 
limit of $C_{6}$ $\rightarrow$ 0, the potential
becomes a hard sphere and $r_{c}$ coincides with the $s$-wave
scattering length $a_{sc}$. For the potential including the long
range part, a tiny change in $r_{c}$ may cause an
enormous change in $a_{sc}$, including sign~\cite{ffd}. As $r_{c}$
decreases from a large value, $a_{sc}$ decreases, and at a particular
critical value of $r_{c}$, it passes through an infinite discontinuity
from $-\infty$ to $+\infty$~\cite{ffd}. Thereafter the potential
supports a two body bound state.  This pattern repeats as $r_{c}$
decreases further.  Positive values of $a_{sc}$ correspond to repulsive 
potential whereas negative $a_{sc}$ values correspond to attractive 
potential. Thus minute tuning of $r_{c}$ can cause the effective potential
to change from attractive to repulsive. 
In the mean-field description, the two-body interaction is solely  represented by the $s$-wave scattering length $a_{sc}$ and 
depending on its sign, positive or negative, the condensate is treated as  repulsive or attractive respectively. However in 
our many-body calculation, we solve zero-energy two-body Schr\"odinger  equation 
with $V(r_{ij})$ given above and tune $r_{c}$ to obtain correct value 
of $a_{sc}$ which mimics the $^{7}$Li condensate of Rice University~\cite{rrf}.  However with a tiny change in $r_{c}$, $a_{sc}$ may change by large amount  including the sign. Each 
additional sign change means that the potential will support an extra 
two-body bound state and it results in an extra node in 
$\eta(r_{ij})$. We choose $r_{c}$, such that it corresponds to the zero  node in the two-body wave function. 
The chosen parameter for our calculation is 
$C_{6}$ = 1.71487 $\times$ 10${^{-12}}$ o.u. and 
$r_{c}$ = 5.3378 $\times 10^{-4}$ o.u.. 
With these sets of parameters we solve the set of coupled differential  equations by the hyperspherical adiabatic approximation (HAA)~\cite{das2}. 
In HAA, one assumes that the hyperradial motion is slow compared to 
the hyperangular motion. The effective potential for the hyperradial 
motion (obtained by diagonalizing the potential matrix together with the  diagonal hypercentrifugal repulsion for each value of $r$) 
is obtained as a parametric function of $r$. We choose the lowest
eigenpotential ($\omega_0(r)$) as the effective potential.
Thus in HAA, energy and wavefunction are obtained approximately by solving a single uncoupled differential equation 
\begin{equation}
\left[ -\frac{\hbar^{2}}{m} \frac{d^{2}}{dr^{2}} + \omega_{0}(r)+
\sum_{K=0}^{K_{max}} |\frac{d\chi_{K0}(r)}{dr}|^{2}-E \right]
\zeta_{0}(r)=0,
\end{equation}
subject to appropriate boundary conditions on $\zeta_{0}(r)$. 
The third term is a correction to the lowest order HAA 
approximation. $\chi_{K0}(r)$ is the $K$-th component ($K$ being 
the hyperangular momentum quantum number) of the eigenvector 
corresponding to the lowest eigenvalue ($\omega_0(r)$) of potential 
plus hypercentrifugal matrix.  
This is called uncoupled adiabatic approximation (UAA), whereas 
disregarding the third term corresponds to the extreme adiabatic  approximation (EAA).
The principal advantages of the present method are as follows. 
First, the correlated PH basis keeps all possible two-body 
correlations and the number of variables is reduced to only 
four irrespective of the number of bosons in the trap. So by 
this method, we can treat quite a large number of atoms without 
much computational difficulty.
Second, the use of HAA basically reduces the multidimensional 
problem to an effective one-dimensional one introducing the effective 
potential. The effective potential $(\omega_{0})$ gives clear 
qualitative as well as quantitative pictures. For our numerical 
calculation we fix 
$l=0$ and truncate the CPH basis to a maximum value $K$ = $K_{max}$,  requiring proper convergence. 
Third, the use of van der Waals interaction with a finite range 
is more realistic than the use of a zero-range contact 
interaction, as in the mean-field GP approach. The pathological 
singularity of the $\delta$-function attractive potential does 
not arise in the present treatment. \\

In Fig.~1, we plot the many-body effective 
potential $\omega_{0}(r)$ as a function of hyperradius $r$ for 
$N$=200 atoms in the condensate. 
As $N=200$ is less than $N_{cr} \simeq 1300$~\cite{rrf}, 
the condensate is metastable and is associated with a deep 
and narrow attractive well (NAW) on the left side. 
For $r \rightarrow 0$, there is a strong repulsive wall. 
This is the immediate reflection of using 
hard core van der Waals interaction. 
The effective potential for an attractive contact interaction 
goes rapidly to $-\infty$ as $r \rightarrow 0$, causing an 
essential and pathological singularity. 
Thus, our many-body picture with nonlocal interaction is in 
sharp contrast with the GP 
mean-field picture with local interaction. The nonlocal interaction 
and the repulsive core of the van der Waals potential 
prevent the Hamiltonian from being unbound from below. 
For $N$ less than the critical value $N_{cr}$, a metastable 
region (MSR) appears for larger $r$, an intermediate barrier (IB) 
separating the NAW and the MSR. For still larger $r$, the influence of the 
attractive interaction subsides and the external wall of the harmonic 
trap dominates. 
In panel~(a) of Fig.~1, the NAW together with the repulsive core is 
shown. In panel~(b) of the same figure, the IB and MSR have been 
plotted. The bottom of the NAW has a very large magnitude compared 
with the bottom of the MSR, hence they cannot be shown in the same 
figure. Note the widely different scales used in the two panels. 
Furthermore, $r$ in panel~(b) is in logarithmic scale.\\

With the increase in the number $N$, we 
observe a decrease in the height of the intermediate barrier, 
together with a decrease in the difference between the maximum of 
IB and the minimum of MSR 
and the NAW starts to be more negative and narrower. 
As $N \rightarrow N_{cr}$, the maximum of IB and the minimum 
of MSR merge to form a point of inflexion, with the disappearance of 
the MSR. At $N=N_{cr}$, the metastable condensate collapses. For 
$N \geq N_{cr}$, there will be only the NAW and no metastable condensate. 
In our present study we observe the IB just vanishes and the 
condensate collapses at $N= 1430$. This is the usual collapse of the 
attractive condensate. At this point, all the atoms 
get trapped in the NAW and form van der Waals cluster which 
corresponds to a high-density branch in the density profile. \\

Next we study different correlation properties in such a 
realistic condensate. We also investigate how the coherence 
properties 
depend on the trap size ($a_{ho}$) and determine the critical 
size of the trap. \\

\section*{B : One-body density}
It was originally pointed out that the BEC is evidenced by the 
presence of off-diagonal long-range order in the one-particle 
density matrix~\cite{pen}. However this definition is not strictly valid 
in the case of a finite-size inhomogeneous system. Due to the 
presence of an external harmonic trap, our system is inhomogeneous. 
In our many-body formalism, we define the one-body density as the 
probability density of finding a boson at a distance $\vec{r}_k$ 
from the center of mass of the condensate as 
\begin{equation}
 R_1(\vec{r}_k)=\int_{\tau'} |\psi|^2 d\tau'.
\end{equation}
where $\psi$ is the full many-body wave function and the integral over the hypervolume $\tau'$ excludes the variable 
$\vec{r}_k$. The incremental hypervolume $d\tau'$ is given by 
\begin{equation}
 d\tau' = r'^{3N-4} \cos^2 \phi \sin^{3N-7}\phi dr' d\phi d\omega_{ij} d\Omega_{N-2}
\end{equation}
where $r'$ is obtained from the relation 
\begin{equation}
 r^2 = r'^{2} + 2 r_k^2
\end{equation}
and the other symbols have their usual meanings~\cite{was}. The 
integral is computed analytically followed by numerical 
computation using a 32-point Gaussian quadrature with the 
original interval divided into progressively increasing 
subintervals. According to Penrose and Onsager definition 
of Bose-Einstein condensation, 
the one-particle density matrix must be associated with a 
single macroscopic eigenvalue~\cite{pen}. However the function 
$\psi(r)$ can be directly expressed in terms of first-order 
correlation function~\cite{nar} and all the information of the one-body density correlation 
are contained in Eq.~(17). The short-range repulsion 
between interacting atoms give some new aspects in the atomic 
correlation. We present our results in Fig.~2 for 200 $^7$Li 
atoms with $a_{sc}=-27.3$~Bohr and two different trap sizes: trap 
frequency $\omega=1.01$ kHz (corresponding to trap size 
$a_{ho}=3.0$ $\mu$m) and $\omega=2.27$ kHz (corresponding to trap size 
$a_{ho}=2.0$ $\mu$m). The one-body density profile deviates from the Gaussian profile of non-interacting case. For comparison, we 
include the mean-field GP results. The deviation from the 
GP result is attributed to the effect of interatomic correlation. 
For the reduced trap size of $2.0$ $\mu$m, we observe that the 
peak becomes higher and narrower, the difference between the mean-field GP and many-body result 
also increases, as the system develops shorter range correlations. 
This can be understood from the fact that in the standard GP approach 
the NAW is disregarded, which pulls the system inwards. 
The many-body results 
for gradually reduced trap sizes are presented in Fig.~3. It was 
observed earlier that for the homogeneous system the long-range 
behavior in the one-body density follows a power-law decay~\cite{nar}. 
However in the presence of an external trap the 
long-range tail in the one-body density is of the order of the trap 
width. Thus by reducing the trap size gradually we observe that 
the long tail in the one-body density decreases as expected. 
For a very tight trap, the one-body density is sharply 
peaked and the long-range order is sharply reduced, in tune with 
reduced trap size. It indicates 
that the atoms in the metastable condensate are highly correlated. 
%and in the extreme case of 
%a very tight trap it will exhibit a $\delta$-type peak. It 
%implies the destruction of the metastable condensate due to the 
%strong interatomic correlations. 
These features become 
more clear from Fig.~4, where we plot the change in the MSR 
of the effective many-body potential with the change in the trap 
size. In Table~I we present the 
position and the value of the maximum of IB and the second 
minimum (which comprises the MSR) in the column~3 and 4 respectively. 
Although the position and the value of the maximum of IB do 
not change much with 
the trap size, the position and the value of second minimum 
are greatly shifted. It indicates that the MSR is pulled in 
and the condensate shrinks with decrease in trap size. 
The corresponding interaction 
energy is presented in column~5 of Table~I. With decrease in 
trap size the attractive interaction energy sharply increases 
and we find $a_{ho}=0.42$ $\mu$m  
as the critical trap size. Just below this critical size the 
condensate will be 
destroyed due to high quantum fluctuations as the negative 
interaction energy increases very fast for smaller $a_{ho}$. 
What actually happens is the following. As the trap size is 
reduced, the system gets squeezed. Consequently, the 
interatomic spacing reduces and the net attractive 
interaction energy increases. Eventually, when the 
interatomic spacing reaches the typical cluster size, 
tightly bound clusters are formed, together with the complete 
removal of the metastable condensate. Such a scenario 
is not possible with an attractive contact interaction, 
for which the effective potential goes to $-\infty$ as 
$r \rightarrow 0$. 
Note that this collapse is different 
from the usual collapse for $a_{ho}=3 \mu$m, where the 
metastable region 
vanishes at $N = N_{cr}$. The critical trap size strongly depends 
on the number of bosons in the trap, which may be just few 
tens to a few hundreds depending on the trap frequency.  
For $N=200$ atoms the critical trap frequency is 
$\omega = 51.47$ kHz. \\

The healing length is sometimes referred to as the coherence length 
and may be considered a relevant quantity to quantify the 
correlation in very 
tight traps. Healing length $\zeta$ is basically the minimum 
distance over which the order parameter can heal. It is in 
general calculated by balancing the quantum pressure and the 
interaction energy of the condensate. In Fig.~5 we plot the 
healing length as a function of the 
trap size, which shows a steep decrease in $\zeta$ in very 
tight traps. The simple expression for healing length is 
$\zeta  =  \frac{1}{\sqrt{8 \pi a_{sc}n(0)}}$, where $n(0)$ is 
the peak density at $r=0$. However in principle $n$ depends 
on $r$ and $\zeta(r)  =  \frac{1}{\sqrt{8\pi a_{sc}n(r)}}$, i.e., 
as $n(r) \rightarrow 0$, $\zeta(r)\rightarrow \infty$ at the 
surface of the cloud.
We can also compare our results with the 
simple Gaussian estimate, for which  
$n(r) = \frac{N}{(\sqrt{\pi a_{ho})^{3}}}e^{-\frac{r^{2}}{a_{ho}^{2}}}$ 
and $\zeta(0) \simeq a_{ho}^{3/2}\sqrt{8\pi a_{sc}N}$. 
Thus $\zeta$ scales as $a_{ho}^{3/2}$. Thus the smooth decrease of the healing length in Fig. 5 reflects the 
effect of interatomic interaction and two-body correlation.\\

As mentioned earlier, in the standard GP mean-field theory, 
the metastable region vanishes as $N \rightarrow N_{cr}$ and 
the condensate collapses into the singular well. The fate of 
the condensate is not predicted further. Here, the many-body 
picture is different. In our case the 
narrow attractive well on the left facilitates further 
study of the condensate near the criticality. The metastable 
condensate leaks through the intermediate 
barrier, and settles down in the NAW. The atoms in the 
condensate form van der Walls clusters via strongly enhanced 
three- and higher-body collisions. It corresponds to the 
high-density branch~\cite{xxz}. 
In Fig.~6 we demonstrate graphically the dependence of the NAW on 
the trap size. In Table~I, we present the position and value of 
the deep minimum of the NAW in column~1. The position of the 
deep well and the adjacent barrier, which comprise the NAW, do 
not change with the trap size. These clearly demonstrate that 
the high-density branch remains invariant with the trap size. The 
calculated size of the atomic cluster is of the order of $0.005$ $\mu$m. \\

This high-density stable state within the NAW corresponds to the Li clusters. 
It would be interesting if these clusters could be detected experimentally. 
Although there is no experimental study of $^{7}Li_{N}$ clusters 
till now, bosonic 
$^{4}He_{N}$ clusters have been studied experimentally. The size of $^{4}$He clusters is determined by the matter 
wave diffraction gratings ~\cite{rbu}. This is a very promising technique for the study of $^{4}$He$_{N}$ clusters. 
However, such techniques may not be feasible for clusters formed from 
collapsed BEC, due to small number of such clusters and difficulty in extracting a beam for the diffraction experiment. Perhaps a better 
method may be using spectroscopic techniques~\cite{yxu}. 
%A similar type of experiment may be feasible for the study of $^{7}$Li 
%clusters in future experiments. 
Utilizing the Feshbach resonance the effective interaction between two atoms can be changed essentially to any value as 
desired and it facilitates the creation of large weakly bound clusters. The signature of the universal behavior of weakly bound bosonic cluster is 
studied in Ref.~\cite{blue} and can be observed in ultracold Bose gas.

\section*{C: Pair-distribution ~function}
So far we have focused on the one-particle density which 
basically considers analysis of the order parameter and 
contains all information about the 
one-particle aspects. In the present section we 
are interested to calculate the probability 
$R_{2}(r_{ij})$ which is defined as the probability of 
finding the $(ij)$-pair of particles at a relative separation 
$r_{ij}$. 
For a correlated interacting system we define 
$R_{2}(r_{ij})$ as follows
\begin{equation}
R_2(r_{ij})=\int_{\tau ''}|\psi|^2d\tau '',
% R_{2}(r_{ij})=\int_{\tau'}|\psi_{ij}(\vec{r}_{ij},\rho_{ij},
%\Omega^{(N-1)})|^{2} d\tau'
\end{equation}
where $\psi$ is the many-body wavefunction as before but now the integral over the hypervolume $\tau ''$ excludes integration over 
${r}_{ij}$. The incremental hypervolume $d\tau ''$ is given by
\begin{equation}
 d\tau'' = r_{ij}^2 (r\sin\phi)^{3N-4} d\omega_{ij} 
d\rho_{ij} d\Omega_{N-1}
\end{equation}
Our results for the same number of $^7$Li atoms and for the same 
trap sizes as in Fig.~3 are presented in Fig.~7. 
When $R_{2}(r_{ij})$ =0, there is no diagonal correlation. The 
peak value of $R_{2}$ at some distance $r_{ij}$ signifies strong 
clustering effects. 
The study of pair correlation is important as the realistic 
interatomic interaction with a small hard core repulsion plays 
a crucial role. It forbids the atoms to come {\it too close} 
due to nucleus-nucleus repulsion. 
Thus $R_{2}(r_{ij})$ is always zero at $r_{ij}=0$. 
Unlike the uniform system with no confinement, our calculation 
shows that $R_{2}$ vanishes asymptotically. 
This is the effect of external confinement 
which restricts the pair separation to a finite value. 
In our earlier calculations we observed the 
dependence of correlation function on the interaction 
strength~\cite{was}. However here we observe 
that even for weak interaction and just a few hundred 
atoms, correlation length decreases drastically in 
a tight trap. To be more 
quantitative, we define the correlation length as the width of 
correlation function at the half of its maximum value and 
plot it as a function of trap size in Fig.~8.  
In Fig.~8 we also plot the position of clustering spot as a 
function of trap size. Clustering spot is defined as the 
position of $r_{ij}$ where pair-distribution 
function attains a maximum value. Even for such weak interaction, 
we observe that both the correlation length and position of 
cluster spot in the matastable condensate decrease drastically, 
in tune with the trap size, as the trap becomes  
tighter. During this squeezing process, the correlation length 
remains larger than the trap size. It indicates that the 
metastable condensate remains highly 
correlated in the tight confinement. But when it is squeezed beyond 
the critical trap size, clustering of particles occurs with the 
destruction of metastable BEC. 

\section{Conclusion}
In an attractive BEC, the destruction of the condensate 
takes place when the number of bosons in the external trap exceeds 
the critical number. At this point, the interaction 
energy becomes large negative (so that the kinetic pressure 
fails to balance it) and the condensate shrinks, such that 
the density at the center of the trap becomes very high. 
The ultimate fate is the collapse of the attractive 
BEC. As the system becomes highly correlated, it is appropriate 
to undertake a correlated many-body approach to 
describe several correlation properties of a realistic condensate 
in a harmonic confinement. The use of a realistic interatomic 
interaction and the finite size trap give the realistic features 
in one-body density and pair correlation properties. 
We calculate such correlation properties of an attractive BEC, 
using the correlated potential harmonics technique. As the trap 
size is given by $a_{ho}=\sqrt{\frac{\hbar}{m\omega}}$, it is easy 
to change the trap size by controlling the laser frequency. A high 
trapping frequency leads to the creation of a very tight trap. 
In this report, we undertake the study of correlation properties 
in a tight trap. Our realistic approach produces an effective potential 
in which the condensate moves. This effective potential for a 
metastable condensate has a 
strongly repulsive core, followed successively by a narrow 
attractive well, an intermediate barrier, a metastable region 
in which the metastable condensate resides and finally the high 
outer wall of the trap. This is in sharp contrast with the 
mean-field GP approach, for which there is an attractive singularity 
on the left of the intermediate barrier. The present approach 
produces a realistic post-collapse scenario. 
We observe that in a very tight trap, the system 
becomes highly correlated and effective correlation length is 
drastically reduced. Even for a small number of atoms well 
below the critical number, the metastable condensate collapses 
when the trap size is below a certain value. One can define a 
critical trap size and critical trap frequency for the existence of a 
metastable BEC in tight confinement for a given number of atoms 
with attractive interaction. Below the critical size, the metastable 
condensate will be destroyed due to high spatial coherence and all 
the atoms will settle down in the narrow attractive well, with the 
formation of coherent van der Waals clusters. For small trap sizes 
above the critical value, the narrow attractive well shows a 
universal character, becoming independent of the trap size. This 
shows that the clusters cannot be further squeezed after their 
formation.
\begin{center}
{\large{\bf{Acknowledgements}}}
\end{center}
AB acknowledges the CSIR, India for a Senior Research Fellowship (Sanction no.: 09/028(0773)/2010-EMR-I).

%\end{document}
%Fig.1
\begin{figure}[hbpt]
\vspace{-10pt}
\centerline{
\hspace{-3.3mm}
\rotatebox{0}{\epsfxsize=8.5cm\epsfbox{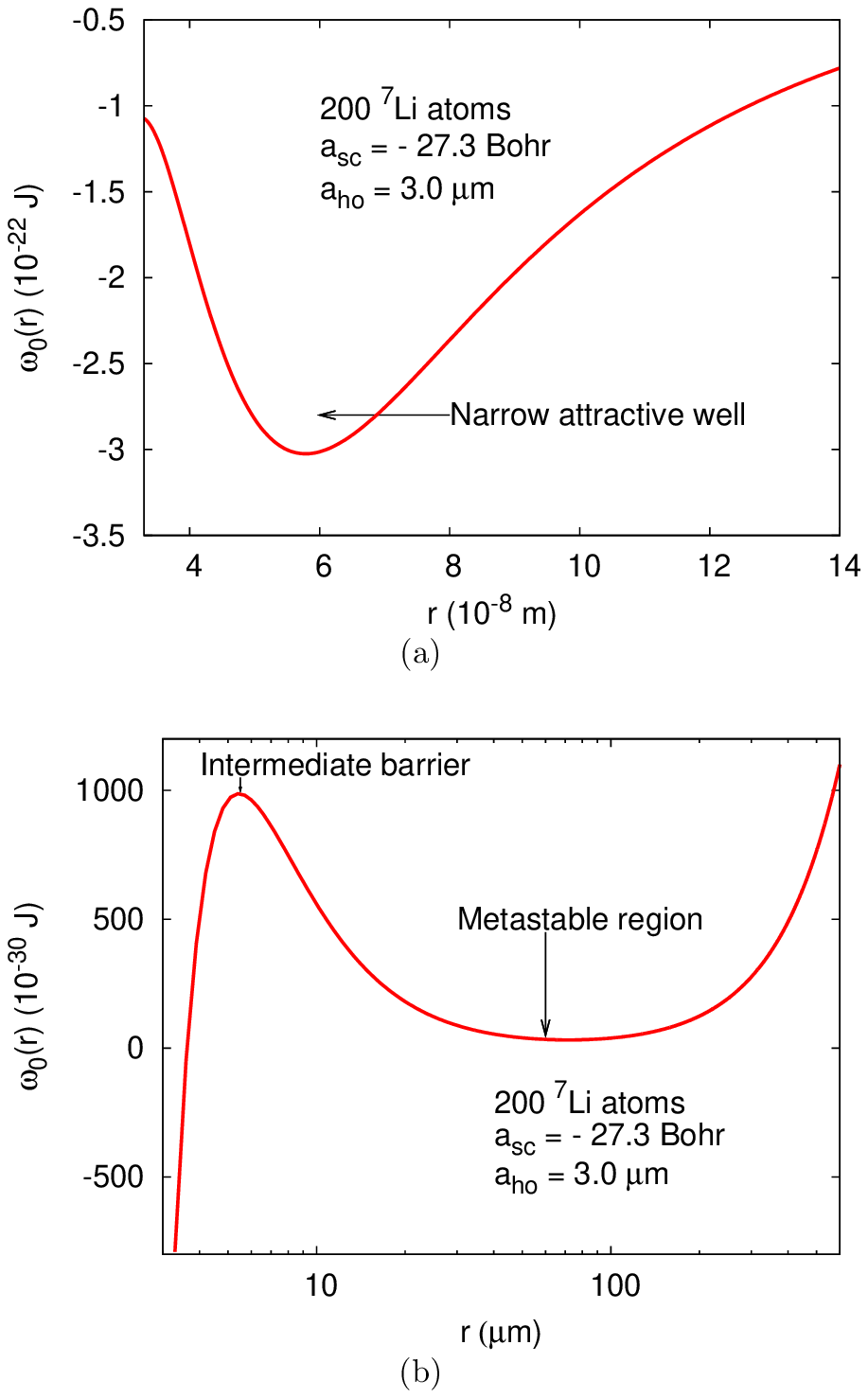}}}
\caption{(Color online) Plot of the effective potential $\omega_{0}(r)$ against $r$ for 200 $^{7}$Li atoms in the 
usual trap of size $a_{ho}$ = 3.0 $\mu m$. The upper panel shows the narrow attractive well (NAW). The lower panel 
shows the metastable region separated from NAW by the intermediate barrier (IB). Note the different scales used for $r$ and $\omega_0(r)$ in the 
two panels, so that NAW is far to the left of the plot in panel~(b) in 
which $r$ is in logarithmic scale. Therefore, the NAW is not 
visible in panel~(b). Note also that the bottom of the NAW would have a value $\approx -3\times 10^{8}$ in the unit used in panel~(b).}
\end{figure}
%=====================================
%   Table 1 ==========================
\begin{table}[!h]
\caption{Parameters for NAW, IB and MSR of the effective potential and the interaction energy $<V>$ for different trap sizes. 
$r_{i}$ and $\omega_{i}$ correspond to the position and value of the 
extrema of NAW ($i=1$), IB ($i=2$) and MSR ($i=3$). Note that different 
units have been used in different columns.}
{\scriptsize{
\begin{center}
\begin{tabular}{|c|c|c|c|c|}
\hline
$a_{ho}$ & NAW & IB  & MSR & $<V>$\\
\cline{2-4}
$~$ & $r_{1}$~~~$\omega_{1}$ & $r_{2}$~~~$\omega_{2}$&$r_{3}$~~~$\omega_{3}$ & $~~$\\
 ($\mu$m) & ($10^{-8}$m)~~~($10^{-22}J$) & ($\mu$m)~~~($10^{-28}J$) & ($\mu$m)~~~($10^{-28}J$) 
& ($10^{-30}J$)\\
\hline
3.0 & 5.79~~~-2.93 & 5.40~~~9.56   & 71.70~~~0.31 & -0.861 \\ 
2.0 & 5.78~~~-2.92 & 5.40~~~9.54   & 47.40~~~0.68 & -2.915 \\ 
1.0 & 5.80~~~-2.92 & 5.50~~~9.61   & 22.80~~~2.61 &  -26. 127\\ 
0.5 & 5.80~~~-2.92 & 5.70~~~10.72  & 10.05~~~9.22 & -306.84\\ 
0.42 & 5.80~~~-2.92 & 6.47~~~12.11 & 7.22~~~ 12.09 & -1321.94\\
\hline
\end{tabular}
\end{center}
}}
\end{table}

%====================================
\begin{figure}[hbpt]
%Fig.-5
\vspace{-10pt}
\centerline{
\hspace{-3.3mm}
\rotatebox{270}{\epsfxsize=6cm\epsfbox{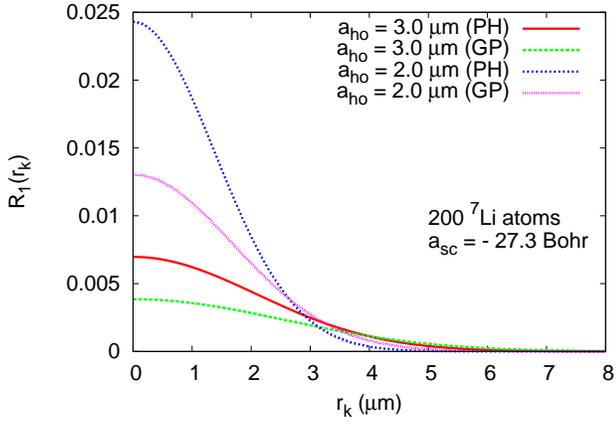}}}
\caption{(Color online) Plot of one-body density ($R_1({r}_k)$) 
against $r_k$ for 200 $^7$Li atoms in the trap with the size of 3.0 $\mu m$ and 2.0 $\mu m$. 
Comparison with the mean field GP results is also presented.}
\end{figure}
%========================
% Fig. 2
\begin{figure}[hbpt]
%Fig.-5
\vspace{-10pt}
\centerline{
\hspace{-3.3mm}
\rotatebox{270}{\epsfxsize=6cm\epsfbox{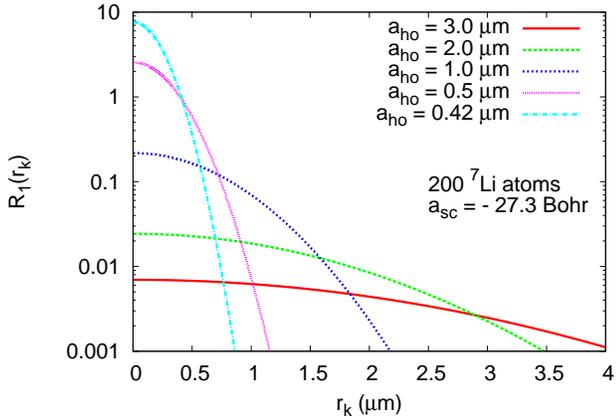}}}
\caption{(Color online) Plot of one-body density ($R_1({r}_k)$) 
against ${r}_k$ for the same number of atoms as before with different trap sizes.}
\end{figure}
%---------------------
% Fig.3
\begin{figure}[hbpt]
%Fig.-5
\vspace{-10pt}
\centerline{
\hspace{-3.3mm}
\rotatebox{270}{\epsfxsize=6cm\epsfbox{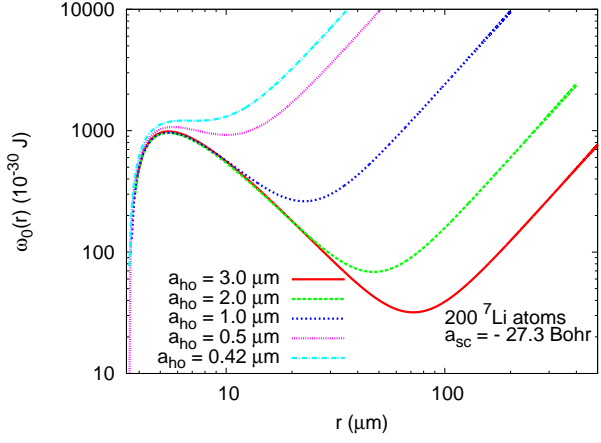}}}
\caption{(Color online) Change in the position and the depth of metastable region for 200 atoms with 
different trap sizes.}
\end{figure}
%==========================
% Fig. 4
\begin{figure}[hbpt]
%Fig.-5
\vspace{-10pt}
\centerline{
\hspace{-3.3mm}
\rotatebox{270}{\epsfxsize=6cm\epsfbox{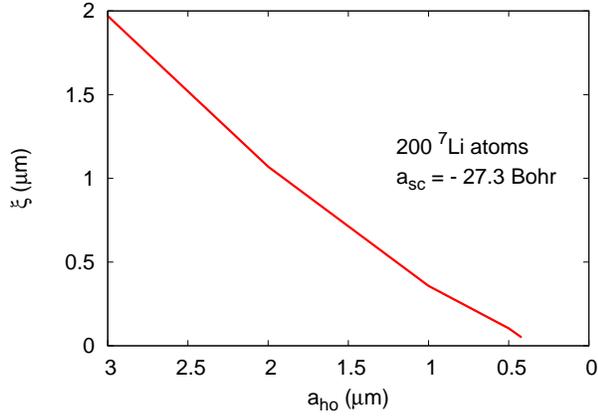}}}
\caption{(Color online) Plot of healing length as a function of trap size.}
\end{figure}
%========================
% Fig. 5
%  Draw it 
\begin{figure}[hbpt]
%Fig.-5
\vspace{-10pt}
\centerline{
\hspace{-3.3mm}
\rotatebox{270}{\epsfxsize=6cm\epsfbox{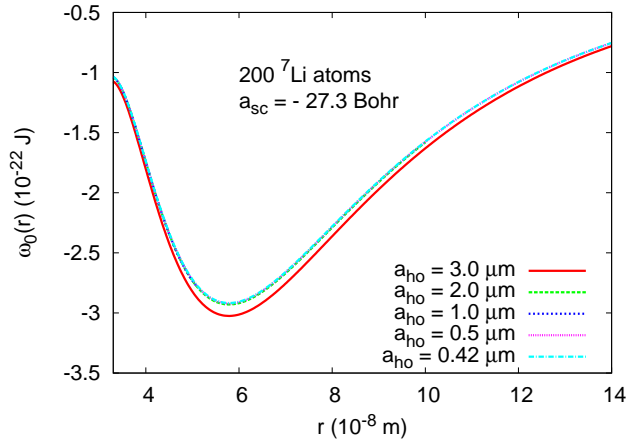}}}
\caption{(Color online) Plot of narrow attractive well for various trap sizes.}
\end{figure}
%=====================
% Fig. 6
\begin{figure}[hbpt]
%Fig.-5
\vspace{-10pt}
\centerline{
\hspace{-3.3mm}
\rotatebox{270}{\epsfxsize=6cm\epsfbox{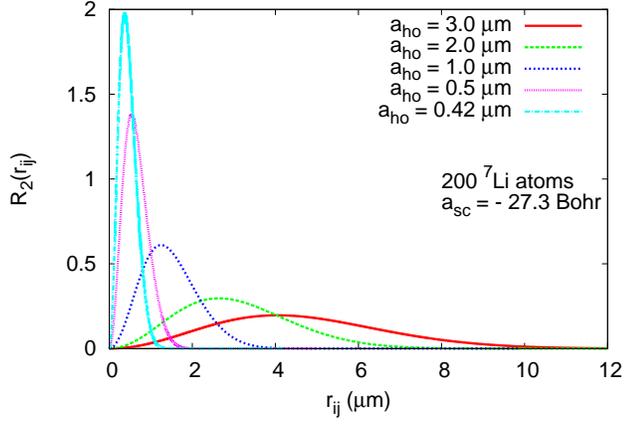}}}
\caption{(Color online) Plot of pair distribution function ($R_2(r_{ij})$) 
against $r_{ij}$ for an attractive Bose 
gas with different trap sizes.}
\end{figure}
%=========================
%  Fig. 7 Draw it
%  Calculate clustering spot as the position where maximum of R_2 occurs.
% You have the value of fwhm.
% Plot them together as a function of trap size.
\begin{figure}[hbpt]
%Fig.-5
\vspace{-10pt}
\centerline{
\hspace{-3.3mm}
\rotatebox{270}{\epsfxsize=6cm\epsfbox{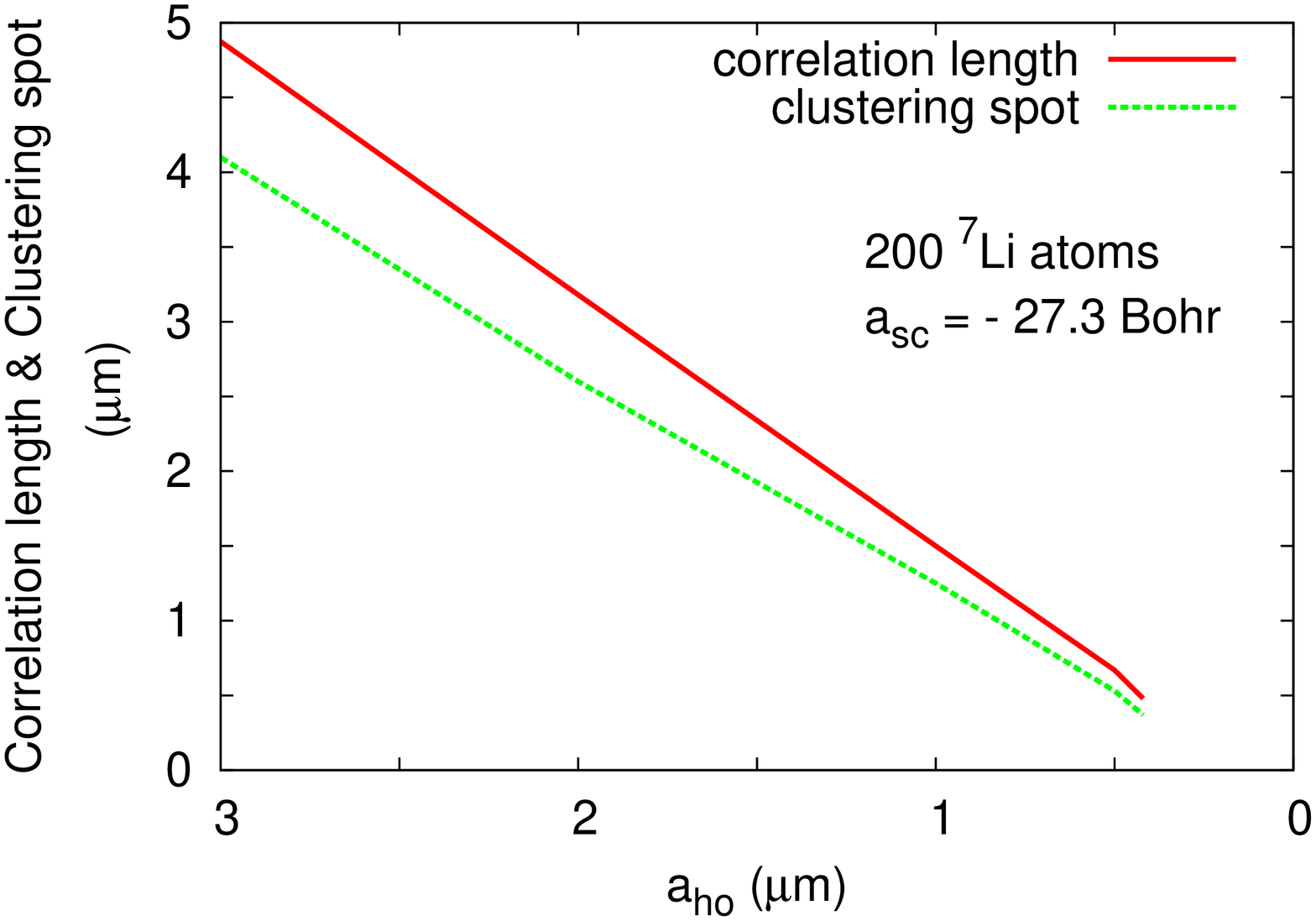}}}
\caption{(Color online) Plot of correlation length and clustering spot as a function of trap size.}
\end{figure}


\begin{thebibliography}{References}
\bibitem{pen} O. Penrose and L. Onsager, {Phys. Rev. A {\bf{104}}, 576 (1956)}.
\bibitem{ujh} C. N. Yang, {Rev. Mod. Phys. {\bf{34}}, 694 (1962)}.
%\bibitem{sak} K. Sakman {\it{et al}}, {Phys. Rev. A {\bf{78}}, 023615 (2008)};
% L. E. Sadler {\it{et al}}, {Phys. Rev. Lett. {\bf{98}}, 110401 (2007)}.
%\bibitem{ast} G. E. Astrakharchik and S. Giorgini, {Phys. Rev. A {\bf{68}}, 031602(R) (2003)}
%\bibitem{ner} S.Z\"ollner, H.D.Meyer and P.Schmelcher, {Phys. Rev. Lett. {\bf{100}}, 040401 (2008)}
%\bibitem{ijy} R. Pezer and H. Buljan, {Phys. Rev. Lett. {\bf{98}}, 240403 (2007)}
%\bibitem{khr} K. V. Kheruntsyan, {Phys. Rev. Lett. {\bf{91}}, 040403 (2003)}
%\bibitem{gan} D. M. Gangardt and G. V. Shlyapnikov, {Phys. Rev. Lett. {\bf{90}}, 010401 (2003)}
%\bibitem{rit} S. Ritter {\it{et al}}, {Phys. Rev. Lett. {\bf{98}}, 090402 (2007)}.
%\bibitem{tgg} I. Bloch, T. H\"ansch and T. Esslinger, {Nature(London) {\bf{403}}, 166 (2000)}.
\bibitem{das} T.K.Das and B.Chakrabarti, {Phys. Rev. A {\bf 70}, 063601, (2004)}.
\bibitem{das1} T.K.Das, S.Canuto, A.Kundu and B.Chakrabarti, {Phys. Rev. A {\bf 75}, 042705, (2007)}.
\bibitem{ffd} T. K. Das, A. Kundu, S. Canuto and B. Chakrabarti, {Phys. Lets A {\bf{373}}, 258 (2009)}. 
\bibitem{Salasnich} L. Salasnich, {Phys. Rev. A {\bf 61}, 015601 (1999)}. 
\bibitem{Parola} A. Parola, L. Salasnich and L. Reatto, {Phys. Rev. (Rapid Comm.) 
{\bf 57}, R3180 (1998)}.
\bibitem{Reatto} L. Reatto, A. Parola and L. Salasnich, {J. Low Temp. Phys. {\bf 113}, 
195 (1998).}
\bibitem{fabre} M.Fabre {\it de la} Ripelle, {Ann. Phys. (N.Y.) {\bf 147}, 281, (1983)};
 J.L.Ballot and M.Fabre {\it de la} Ripelle, {Ann. Phys. (N.Y.) {\bf 127}, 62, (1980)}.
\bibitem{Abramowitz} M. Abramowitz and I. A. Stegun, {\it Handbook of 
Mathematical Functions}, Dover Publications Inc., New York (1972). 
\bibitem{chak1} B. Chakrabarti and T. K. das, Phys. Rev. A {\bf 78}, 063608 
(2008).
\bibitem{debnath} P. K. Debnath, B. Chakrabarti and T. K. Das, Int. J. 
Quan. Chem. {\bf 111}, 1333 (2011). 
\bibitem{haldar} S. K. Haldar, B. Chakrabarti and T. K. Das, Phys. Rev. A 
{\bf 82}, 043616 (2010). 
\bibitem{kundu} A. Kundu, B. Chakrabarti, T. K. Das and S. Canuto, J. Phys. 
B {\bf 40}, 2225 (2007).
\bibitem{rrf} C. C. Bradlay, C. A. Sackett and R.G. Hulet, {Phys. Rev. Lett. {\bf{78}}, 985 (1997)}.
\bibitem{das2} T.K.Das, H.T.Coelho and M.Fabre {\it de la} Ripelle, {Phys. Rev. C {\bf 26}, 2288, (1982)}.
\bibitem{was} A. Biswas, B. Chakrabarti and T. K. Das {J. Chem. Phys. {\bf{133}}, 104502 (2010)}.
\bibitem{nar} M. Naraschewski and R.J.Glauber, {Phys. Rev. A {\bf{59}}, 4595 (1999)}
\bibitem{xxz} A. Biswas, T. K. Das, L. Salasnich and B. Chakrabarti {Phys. Rev. A {\bf{82}}, 043607 (2010)}.
\bibitem{rbu} R. Br\"uhl {\it et al}, {Phys. Rev. Lett. {\bf{95}}, 063002 (2005)}. 
\bibitem{yxu} Y. Xu and W. J\"ager, Phys. Chem. Chem. Phys. {\bf 2}, 
3549 (2000). 
\bibitem{blue} G. J. Hanna and D. Blume, {Phys. Rev. A {\bf{74}}, 063604 (2006)}.
\end{thebibliography}
\end{document}